# Mesoscopic Model for Mechanical Characterization of Biological Protein Materials

Gwonchan Yoon[1], Hyeong-Jin Park[1], Sungsoo Na[1,*], and Kilho Eom[2,†]

[1]*Department of Mechanical Engineering, Korea University, Seoul 136-701, Republic of Korea*

[2]*Nano-Bio Research Center, Korea Institute of Science & Technology (KIST), Seoul 136-791, Republic of Korea*

---

[*] Corresponding Author. E-mail: nass@korea.ac.kr
[†] Corresponding Author. E-mail: eomkh@kist.re.kr

**Abstract**

Mechanical characterization of protein molecules has played a role on gaining insight into the biological functions of proteins, since some proteins perform the mechanical function. Here, we present the mesoscopic model of biological protein materials composed of protein crystals prescribed by Go potential for characterization of elastic behavior of protein materials. Specifically, we consider the representative volume element (RVE) containing the protein crystals represented by $C_\alpha$ atoms, prescribed by Go potential, with application of constant normal strain to RVE. The stress-strain relationship computed from virial stress theory provides the nonlinear elastic behavior of protein materials and their mechanical properties such as Young's modulus, quantitatively and/or qualitatively comparable to mechanical properties of biological protein materials obtained from experiments and/or atomistic simulations. Further, we discuss the role of native topology on the mechanical properties of protein crystals. It is shown that parallel strands (hydrogen bonds in parallel) enhance the mechanical resilience of protein materials.

*Keywords:* Mechanical Property; Protein Crystal; Go Model; Virial Stress; Young's Modulus

## INTRODUCTION

Several proteins bear the remarkable mechanical properties such as super-elasticity, high yield-strength, and high fracture toughness.[1-5] Such remarkable properties of some proteins have attributed to the mechanical functions. For instance, spider silk proteins exhibit the super-elasticity relevant to spider-silk's function.[4,5] Specifically, the super-elasticity of spider silk plays a role on the ability of spider silk to capture a prey such that high extensibility enables the spider silk to convert the kinetic energy of flying prey into the heat dissipation, resulting in the capability of capturing the prey. Furthermore, it has recently been found that spider silk protein possesses the remarkable mechanical properties such as yield strength comparable to that of high-tensile steel and fracture toughness better than that of Kevlar.[6] This highlights that understanding of mechanical behavior of protein materials such as spider silk may provide the key concept for design of biomimetic materials, and that mechanical characterization of protein materials may allow for gaining insight into the biological functions of mechanical proteins.

Mechanical characterization of biological molecules such as proteins has been successfully implemented by using atomic force microscopy (AFM), optical tweezers, or fluorescence method. AFM has been broadly employed for characterization of mechanical bending motion of nanostructures such as suspended nanowires,[7-9] and biological fibers such as microtubules.[10] Fluorescence method for a cantilevered fibers such as microtubules[11] and/or DNA molecules[12] has allowed one to understand the relationship between persistent length (related to bending rigidity) and contour length, enabling the validation of the continuum model of biomolecules such as microtubule and DNA. In last decade, since the pioneering works by Bustamante and coworkers[13,14] and Gaub and coworkers,[15,16] optical tweezer and/or AFM has enabled them to characterize the microscopic mechanical behavior of proteins such as protein unfolding mechanics. Such protein unfolding experiments has been illuminated in that these studies may provide the free energy landscape of proteins related to protein folding mechanism.[17,18] Nevertheless, microscopic characterization such as protein unfolding mechanics may not be sufficient to understand the remarkable mechanical properties of biological materials.

Computational simulation for mechanical characterization of proteins has been taken into account based on atomistic model such as molecular dynamics[19] and/or coarse-grained model.[20] Atomistic model such as steered molecular dynamics (SMD) simulation has allowed one to gain insight into protein unfolding mechanics.[19,21] However, such SMD simulation has been still computationally limited to small proteins since the time scale available for SMD is not relevant to the time scale for AFM experiments of protein unfolding mechanics. Recently, the coarse-grained model such as Go model has been recently revisited for mimicking the protein unfolding experiments.[20,22] It is remarkable that such revisited Go model has provided the protein unfolding behavior quantitatively comparable to AFM experiments, and that it has also suggested the role of temperature, AFM cantilever stiffness, and other effects on protein unfolding mechanism.[23] Eom *et al*[24,25] provided the coarse-grained model of folded polymer chain molecules for gaining insight into unfolding mechanism with respect to folding topology, and it was shown that folding topology plays a role on the protein unfolding mechanism.

However, the computational simulations aforementioned have been restricted for understanding the microscopic mechanics of protein unfolding. The macroscopic mechanical behavior of protein crystals has not been much highlighted based on computational models, albeit there have been few literatures[26-28] on macroscopic mechanical behavior of protein crystals. Termonia *et al*[29] had first provided the continuum model of spider silk such that their model regards the spider silk as β-sheets connected by amorphous Gaussian chains. Even though such model reproduce the stress-strain relationship for spider silk comparable to experiments, this model may be inappropriate since spider silk has been recently found to consist of β-sheets and ordered α-helices.[30] Zhou *et al*[31] suggested the hierarchical model for spider silk in such a way that spider silk is represented by hierarchical combination of nonlinear elastic springs, inspired by AFM experimental results by Hansma and coworkers.[4] Kasas *et al*[32] had established the continuum model (tube model) for microtubules based on their AFM experimental results. These continuum models and/or hierarchical model mentioned above are phenomenological models for describing the macroscopic mechanical properties of biological materials.

There have been few literatures[26-28] on the characterization of macroscopic mechanical properties such as Young's modulus of biological materials such as protein crystals and fibers based on physical model such as atomistic model (e.g. molecular dynamics simulation) for protein crystal. Despite of the ability of atomistic model to provide the macroscopic properties of protein crystals,[28] the atomistic model has been very computationally restricted to small protein crystals.

In this work, we revisit the Go model in order to characterize the macroscopic mechanical properties of biological protein materials composed of model protein crystals such as α helix, β sheet, α/β tubulin, titin Ig domain, etc. (See Table 1). Specifically, we consider the representative volume element (RVE) containing protein crystals in a given space group for computing the virial stress of RVE in response to applied macroscopic constant strain. It is shown that our mesoscopic model based on Go model has allowed for estimation of the macroscopic mechanical properties such as Young's modulus for protein crystals, quantitatively comparable to experimental results and/or atomistic simulation results. Moreover, our mesoscopic model enables us to understand the structure-property relationship for protein crystals. The role of molecular structure on the macroscopic mechanical properties for protein crystals has also been discussed. It is provided that, from our simulation, the native topology of protein structure is responsible for mechanical properties of protein crystals.

## MODELS

### *MESOSCOPIC MODEL FOR BIOLOGICAL PROTEIN MATERIALS*

We assume that the mechanical response of biological materials (fibers), as shown in Fig. 1, can be represented by periodically repeated unit cell referred to as *representative volume element* (RVE) containing the crystallized proteins with a specific space group. We assume that a unit cell is stretched gradually according to the constant, discrete, macroscopic strain tensor $\Delta\boldsymbol{\varepsilon}^0$, where $\Delta\boldsymbol{\varepsilon}^0 = 0.001$. Here, it is also assumed that the unit cell is stretched slowly enough that the time scale of stretching is much longer than that of thermal motion of a protein structure. This may be regarded as a *quasi-equilibrium stretching experiment*, where thermal effect and rate effect are discarded.[24,33] Once a constant, discrete strain tensor $\Delta\boldsymbol{\varepsilon}^0$ is prescribed to a unit cell containing protein crystal, the displacement vector $\mathbf{u}$ due to strain $\Delta\boldsymbol{\varepsilon}^0$ for a given position vector $\mathbf{r}$ of a protein structure is in the form of

$$\mathbf{u}(\mathbf{r}) = \Delta\boldsymbol{\varepsilon}^0 \cdot \mathbf{r} \tag{1}$$

Accordingly, the position vector $\mathbf{r}^*$ of a protein structure after application of discrete,

constant strain tensor to unit cell becomes $\mathbf{r}^* = \mathbf{r} + \mathbf{u}(\mathbf{r})$. Then, we perform the energy minimization process based on conjugate gradient method to find the equilibrium position $\mathbf{r}_{eq}$ for ensuring the convergence of virial stress,[28,34] i.e. $\partial V/\partial \mathbf{r} = 0$ at $\mathbf{r} = \mathbf{r}_{eq}$, where $V$ is the total energy prescribed to protein structure.

For computing the effective material properties of protein crystal, one has to evaluate the overall stress $\boldsymbol{\sigma}^0$ for a unit cell to contain protein crystal due to applied constant, discrete strain $\Delta\boldsymbol{\varepsilon}^0$. The stress $\boldsymbol{\sigma}(\mathbf{r})$ at a position vector $\mathbf{r}$, which is obtained from application of displacement $\mathbf{u}(\mathbf{r}_0)$ for a given position vector $\mathbf{r}_0$ for a protein crystal and consequently energy minimization process, can be computed from the virial stress theory[35,36]

$$\boldsymbol{\sigma}(\mathbf{r}) = \sum_{i=1}^{N}\left[\frac{1}{2}\sum_{j\neq i}^{N}\mathbf{r}_{ij}\otimes\left(\frac{1}{r_{ij}}\frac{\partial\Phi(r_{ij})}{\partial r_{ij}}\mathbf{r}_{ij}\right)\right]\cdot\boldsymbol{\delta}(\mathbf{r}-\mathbf{r}_i) \tag{2}$$

where $N$ is the total number of atoms for a protein crystal in a unit cell, $\mathbf{r}_{ij} = \mathbf{r}_j - \mathbf{r}_i$ with the position vector of $\mathbf{r}_i$ for an atom $i$ in a unit cell, $\Phi(r_{ij})$ the inter-atomic potential for atoms $i$ and $j$ as a function of distance $r_{ij}$ between these two atoms, $\otimes$ indicates the tensor product, and $\boldsymbol{\delta}(\mathbf{x})$ is the delta impulse function. The overall stress $\boldsymbol{\sigma}^0$ can be easily estimated.

$$\boldsymbol{\sigma}^0 \equiv \frac{1}{V}\int_{\Omega} d^3\mathbf{r}\cdot\boldsymbol{\sigma}(\mathbf{r}) = \frac{1}{2V}\sum_{i=1}^{N}\sum_{j\neq i}^{N}\mathbf{r}_{ij}\otimes\left(\mathbf{r}_{ij}\frac{1}{r_{ij}}\frac{\partial\Phi(r_{ij})}{\partial r_{ij}}\right) \tag{3}$$

Here $V$ is the volume of RVE, and a symbol $\Omega$ in the integration indicates the volume integral with respect to RVE.

The process to obtain the stress-strain relationship for protein materials is summarized as below:

(i) We adopt the initial conformation of a protein crystal as the native

conformation deposited in protein data bank (PDB) for a given protein crystal in a unit cell. Such initial confirmation for a protein crystal is denoted as $\mathbf{r}_0$.

(ii) A discrete, constant strain tensor $\Delta\boldsymbol{\varepsilon}^0$ is applied to a unit cell, so that the displacement field $\mathbf{u}$ for a protein crystal in a unit cell is given by $\mathbf{u}(\mathbf{r}_0) = \Delta\boldsymbol{\varepsilon}^0\cdot\mathbf{r}_0$. The atomic position vector for a protein crystal is, accordingly, $\mathbf{r}^* = \mathbf{r}_0 + \mathbf{u}(\mathbf{r}_0)$

(iii) In general, the position vector $\mathbf{r}^*$ is not in equilibrium state, i.e. $\partial V/\partial\mathbf{r}|_{\mathbf{r}=\mathbf{r}^*} \neq 0$. The equilibrium position vector $\mathbf{r}_{eq}$ is computed based on energy minimization (using conjugate gradient method) for an initially given conformation $\mathbf{r}^*$.

(iv) Compute the overall virial stress $\boldsymbol{\sigma}^0$ using Eq. (3) with an atomic position vector of $\mathbf{r} = \mathbf{r}_{eq}$.

(v) Set the initial conformation $\mathbf{r}_0$ as $\mathbf{r}_{eq}$, i.e. $\mathbf{r}_0 \leftarrow \mathbf{r}_{eq}$.

(vi) Repeat the process (ii) – (v) until a unit cell is stretched up to a prescribed strain.

In general, the stress-strain relationship for protein materials obeys the nonlinear elastic behavior. We employ the tangent modulus as the elastic modulus such that the elastic modulus (Young's modulus) is estimated such as $E = \partial\sigma^0/\partial\varepsilon^0$ at $\varepsilon^0 = 0$,[37,38] where $\varepsilon^0$ is the total strain applied to RVE.

***INTER-ATOMIC POTENTIALS: GO MODEL & ELASTIC NETWORK MODEL***

In last decade, it was shown that protein structures can be represented by $C_\alpha$ atoms with an empirical potential provided by Go and coworkers, referred to as *Go model*.[22,23,39] Go model describes the inter-atomic potential for two $C_\alpha$ atoms $i$ and $j$ in the form of

$$\begin{aligned}\Phi\left(r_{ij}\right) &= \left[\frac{k_1}{2}\left(r_{ij}-r_{ij}^0\right)^2+\frac{k_2}{4}\left(r_{ij}-r_{ij}^0\right)^4\right]\delta_{j,i+1} \\ &+4\psi_0\left[\left(\lambda/r_{ij}\right)^6-\left(\lambda/r_{ij}\right)^{12}\right]\left(1-\delta_{j,i+1}\right)\end{aligned} \tag{4}$$

Here, $k_1$ and $k_2$ are force constants for harmonic potential and quartic potential, respectively, $\psi_0$ is the energy parameter for van der Waal's potential, $\lambda$ is the length scale representing the native contacts, superscript 0 indicates the equilibrium state, and $\delta_{i,j}$ is the Kronecker delta defined as $\delta_{i,j} = 1$ if $i = j$; otherwise $\delta_{i,j} = 0$. Here, we used $k_1$ = 0.15 kcal/mol·Å$^2$, $k_2$ = 15 kcal/mol·Å$^2$, $\psi_0$ = 0.15 kcal/mol, and $\lambda$ = 5 Å.[40] The inter-atomic potential in the form of Eq. (4) consists of potential for backbone chain stretching and the potential for native contacts. Go potential is a versatile model for protein modeling such that Go model enables the computation of conformational fluctuation quantitatively comparable to experimental data and/or atomistic simulation such as molecular dynamics.[39] Moreover, Go model has recently allowed one to understand the protein unfolding mechanics qualitatively comparable to AFM pulling experiments for protein unfolding mechanics.[22,23]

Elastic network model (ENM), firstly suggested by Tirion[41] and later by several research groups,[42-47] regards the protein structure as a harmonic spring network. The inter-atomic potential for ENM is given by

$$\Phi\left(r_{ij}\right)=\frac{K}{2}\left(r_{ij}-r_{ij}^o\right)^2\cdot H\left(r_c-r_{ij}^o\right) \tag{5}$$

Here, $K$ is the force constant for an entropic spring ($K$ = 1 kcal/mol·Å$^2$),[42] $r_c$ is the cut-off distance ($r_c$ = 7.5 Å), and $H(x)$ is Heaviside unit step function defined as $H(x) = 0$ if $x < 0$; otherwise $H(x) = 1$. As shown in Eq. (5), the harmonic potential represents the native contacts defined in such a way that the two $C_\alpha$ atoms $i$ and $j$ are connected by an entropic spring with force constant $K$ if the equilibrium distance $r_{ij}^{0}$ between two $C_\alpha$ atoms $i$ and $j$ is less than the cut-off distance $r_c$.

## RESULTS AND DISCUSSIONS

We take into account the biological materials composed of model protein crystals (shown in Table 1) and their mechanical behaviors. The number of residues for model protein crystals ranges from 20 to ~2000, which are typically computationally ineffective for atomistic simulation such as molecular dynamics for mechanical characterization. For mechanical characterization of protein crystals, the constant volumetric strain $e$ is applied to RVE, in which protein crystal resides.

$$e = \frac{1}{3}\left(\boldsymbol{\varepsilon}_{xx}^{0} + \boldsymbol{\varepsilon}_{yy}^{0} + \boldsymbol{\varepsilon}_{zz}^{0}\right) \equiv \frac{1}{3} Tr\left[\boldsymbol{\varepsilon}^{0}\right] \tag{6}$$

where $Tr[\mathbf{A}]$ is the trace of matrix $\mathbf{A}$, and $\varepsilon_{xx}$ is the normal strain induced by extension in longitudinal direction $x$. Once the overall stress for model protein crystal is computed from Eq. (3), the hydrostatic stress (pressure) $p$ can be estimated such as

$$p = \frac{1}{3}\left(\boldsymbol{\sigma}_{xx} + \boldsymbol{\sigma}_{yy} + \boldsymbol{\sigma}_{zz}\right) \equiv \frac{1}{3} Tr\left[\boldsymbol{\sigma}\right] \tag{7}$$

Here, $\sigma_{xx}$ is the normal stress in the longitudinal direction $x$. The constitutive relation provides the material properties such as Young's modulus $E$ and bulk modulus $M$ such as $p = Me$; and consequently, $M = E/[3(1 - 2\nu)]$, where $\nu$ is the Poisson's ratio.[38]

For mechanical characterization of protein materials, we restrict our simulation to *quasi-equilibrium stretching experiments*,[24] where the thermal effect is disregarded. Thermal effect does also play a role in mechanical behavior of protein materials, since thermal fluctuation at finite temperature assists the bond rupture mechanism, i.e. thermal unfolding behavior.[23,48] However, thermal effect does not change the mechanical unfolding pathway related to native topology of protein.[23,48] Also, the bond rupture force (i.e. a peak force, corresponding to the bond rupture event, in the force-extension curve) as well as force-extension curve are insensitive to temperature change near the room temperature.[23] Moreover, the mechanical behavior of materials is generally dependent on stretching rate.[49] The protein unfolding mechanism depends on the pulling rate such that bond rupture force is determined by stretching rate.[25,34,50] However, such stretching rate effect does not affect the unfolding pathway mechanism responsible for mechanical resilience of protein structure.[24,25] Further, rate effect is generally not a control parameter for AFM bending experiment, which provides the Young's modulus of biological materials such as microtubule.[11] Thus, *quasi-equilibrium stretching experiment*, which discards the thermal effect and the stretching rate effect, is sufficient to understand the role of folding topology in the mechanical behavior of protein materials as well as their mechanical properties such as Young's modulus.

The relation between hydrostatic stress and strain for biological protein materials made of model protein crystals are taken into account with virial stress theory based on Go potential prescribed to protein crystal structure. Based on the relationship between hydrostatic stress and strain, we compute the Young's modulus for protein materials composed of model protein crystals (for details, see Table 1). First, let us consider the tubulin as a model protein crystal and its mechanical properties. Tubulin is renowned as a component for microtubules, which plays a mechanical role in maintaining the cell shape. Our simulation provides that the Young's modulus for biological material consisting of tubulin crystal is $E_{tub}$ = 0.138 GPa, which is comparable to AFM bending experiments of microtubule predicted as $E$ = ~0.1 GPa.[10] It is remarkable that our simulation allows for computation of the material property of microtubule based on the tubulin crystals, which is comparable to AFM experimental results. However, it should be noted that estimated Young's modulus by experiments is very sensitive to experimental environments and/or experimental methods. The Young's modulus of microtubule evaluated as $E_{tub}$ = ~0.1 GPa by using AFM bending experiments[10] is different from that using nondestructive method ($E_{tub}$ = ~2.5 GPa)[51] by an order. Such discrepancy in different experiments may be attributed to the role of fiber length on the persistent length of microtubule related to its bending rigidity (elastic modulus).[11] Also, the other effects such as temperature and solvent may affect the estimation of Young's modulus of biological fibers.[10] Further, for validation of our computational model for biological protein materials consisting of protein crystals, as shown in Fig. 2, we also compare the mechanical behavior of titin Ig domains such as proximal and distal domains. Our simulation suggests that distal domain exhibits the better mechanical resistance than proximal domain (i.e. $E_{prox}$ = 0.187 GPa < $E_{dist}$ = 0.254 GPa), in agreement with experimental result showing that distal domain is stiffer than proximal domain.[52]

Fig. 2 depicts the mechanical resistance of biological materials composed of model protein crystals. As mentioned above, the mechanical property such as Young's modulus estimated from our model is quantitatively and/or qualitatively comparable to experimental results (e.g. microtubule, titin Ig domain). It is remarkable that, in Fig. 2, the Young's modulus for biological materials based on model protein crystals is in the range of 0.1GPa to 1 GPa, in agreement with experimental result that Young's modulus for biological materials made from proteins usually ranges from 1 MPa (e.g. elastin) to 10 GPa (e.g. dragline silk).[53] It is also interesting in that our simulation shows that β-sheet exhibits the excellent mechanical resistance such as Young's modulus $E$ and maximum hydrostatic stress, $\sigma_{max}$, among model protein crystals. This is in agreement with previous studies[24,25,34,54] which reported that β-sheet structural motif plays a vital role on toughening the biological materials.

For further understanding the role of molecular interactions as well as topology of protein crystal, we employ the elastic network model (ENM)[41,42] instead of Go potential for computing the virial stress for model protein crystals – α-helix and β-sheet. Since ENM assumes the harmonic potential field to protein structure, the simulation based on ENM predicts the piecewise linear elastic behavior of two model protein crystals. As shown in Fig. 3, the ENM-based simulation overestimates the Young's modulus of two model protein crystals, which may be attributed to the harmonic potential field prescribed to protein structure. This indicates that, for precise quantification of material properties of protein crystal, anharmonic potential field (e.g. Go potential) is necessary. However, it is remarkable that even ENM-based mesoscopic model provides the mechanical resistance of two model protein crystals, qualitatively comparable to our model based on Go potential. Specifically, mesoscopic model based on both ENM and Go model (Go potential) provide that β-sheet possesses the higher Young's modulus than α-helix by factor of ~2. This implies that the material property such as Young's modulus for biological protein material may be correlated with native topology of protein crystal. Moreover, we also consider the fibronectin III (fn3) domains with different crystal structures for understanding the role of protein topology on the material property. As shown in Table 1, our mesoscopic model provides that fn3 domain with a space group of $P4_32_12$ exhibits the higher Young's modulus than those of space groups such as $P2_1$ and/or I2 2 2. This indicates that the topology of crystal structure dictated by space group does also play a role on Young's modulus of protein materials.

In order to gain insight into the role of native topology on the mechanical properties of biological protein materials, we introduce the dimensionless quantity $Q$ representing the degree of folding topology of proteins. For a protein with $N$ residues, the degree-of-fold, $Q$, is defined as $Q = N_c/(N(N-1)/2)$, where $N_c$ is the number of native contacts and $N(N-1)/2$ is the maximum possible number of native contacts. Here, the native contact is defined in such a way that, if two residues are within a cut-off distance (7.5 Å), then these two residues are in the native contact. The degree-of-fold ($Q$) is almost identical to contact-order ($CO$), which is typically used to represent the native topology of proteins (see Fig. 4). Herein, the contact-order is defined such as[55]

$$CO = \frac{1}{L \cdot N_c} \sum \Delta S_{ij} \tag{8}$$

where $L$ is the total number of residues, $N_c$ is the total number of native contacts, and $\Delta S_{ij}$ is the sequence separation, in residues, between contacting residues $i$ and $j$. In Fig. 5, it is shown that the degree-of-fold, $Q$, is highly correlated with Young's modulus, implying the role of contact-order on the Young's modulus for protein materials. Specifically, α-helix and β-sheet exhibit the high degree-of-fold, $Q$, as well as high Young's modulus. On the other hand, some protein materials such as titin Ig domains and TTR have the low degree-of-fold, $Q$, but intermediate value of Young's modulus. This may be ascribed to the fact that titin Ig domain and TTR are known as mechanical proteins which performs the excellent mechanical role due to hydrogen bonding of β-sheet structural motif. This indicates that hydrogen bonding of β-sheet motif plays a significant role in mechanical properties of biological protein materials. Moreover, we also consider the relationship between degree-of-fold, $Q$, and maximum hydrostatic stress, $\sigma_{max}$. As shown in Fig. 6, β-sheet possesses the high degree-of-fold, $Q$, as well as high maximum stress, $\sigma_{max}$, while α-helix exhibits the relatively high degree-of-fold, $Q$, but low maximum stress, $\sigma_{max}$. This may be attributed to the fact that α-helix behaves like a nonlinear helical spring, whereas β-sheet acts like a spring with breakage of hydrogen bonds. In general, the mechanical strength of protein materials is typically originated from the unfolding of folded domain induced by breakage of hydrogen bond.[24,34,54] This is consistent with our simulation results showing that titin Ig domain and TTR have the relatively high maximum stress, $\sigma_{max}$, albeit these protein materials have the low degree-of-fold, $Q$. In other words, the high maximum stress for titin Ig domain and TTR is originated from the β-sheet structural motif that undergoes the bond-breakage upon mechanical loading. This indicates that the β-sheet, which has the high degree-of-fold, $Q$, is responsible for high yield stress of biological protein materials through breakage of hydrogen bond of β-sheet structural motif.

For deeper understanding the role of native topology on the elastic resilience of protein materials, let us consider the polymer chain with hydrogen bonds that can be unfolded in response to external mechanical loading (see Fig. 7). Here, we take into account the two limiting cases: (i) a polymer chain with $N_B$ serial bonds, and (ii) a polymer chain with $N_B$ parallel bonds. For a single bond, the rate for bond-breakage is given by Bell such as $k(f) = k_0\exp(f/f_c)$, where $k(f)$ is the unfolding rate as a function of force ($f$) applied to a single hydrogen bond, and $f_c$ is given as $f_c = k_BT/x_b$ with Boltzmann's constant $k_B$, temperature $T$, and pulling distance $x_b$.[25,54,56-58] The probability for a bond to withstand a force $f$ with a loading rate $\mu$ is $P(f) = \exp[(k_0f_c/\mu)\{1 - \exp(f/f_c)\}]$. Now, consider the case (i) where a polymer chain with $N_B$ serial bonds is pulled with a loading $F$ and a loading rate $\mu$. For this case, the force exerted on every bond is identical such as $f = F$ for every bond. The probability for every bond in serial configuration to be intact under the mechanical loading $F$ is in the form of

$$P(F) = \left[ \exp\left\{ \frac{k_0 f_c}{\mu} \left(1 - \exp(F / f_c)\right) \right\} \right]^{N_B} \tag{8}$$

The probability density $\rho(F)$ to find the first fracture event of any single bond under the mechanical loading $F$ is given by $\rho(F) = -dP/dF$. The most probable mechanical loading, $F_m$, for the first fracture event is obtained from $d\rho/dF = 0$ such as $F_m = f_c \ln[\mu/N_B k_0 f_c]$. This indicates that, for a serial bond, the force at fracture event of a bond has the weak, logarithmic dependence on number of serial bonds. On the other hand, for the case (ii) where parallel bonds reside in the polymer chain, the force exerted for each bond in parallel configuration is given by $f = F/N_B$. The probability to withstand the force $F$ for every bond in parallel is given as

$$P(F) = \left[ \exp\left\{ \frac{k_0 f_c}{\mu} \left(1 - \exp(F / N_B f_c)\right) \right\} \right]^{N_B} \tag{9}$$

In the similar argument to case (i), the most probable mechanical force, $F_m$, for the first fracture event for any single bond is estimated such as $F_m = N_B f_c \ln[\mu/N_B k_0 f_c]$. This suggests that the force corresponding to the rupture of any single bond in parallel configuration is dependent on the number of bonds, $N_B$, with a scaling of $F_m \sim N_B \ln(1/N_B)$. It indicates that the bonds in parallel configuration improve the mechanical resistance to mechanical loading. Conclusively, from these two limiting cases, the configuration of bonds related to native topology of protein structure plays a dominant role on the mechanical resilience dictated by mechanical loading for fracture event of a bond. In other words, the mechanical resilience of proteins is correlated with the native topology characterized by the secondary structure contents, that is, contact order.[55,59-61]

As stated earlier, the bonds in parallel configuration enhances the mechanical resilience, consistent with previous studies[24,34,54] showing that parallel strands in β-sheet structural motif are responsible for mechanical strength of mechanical proteins.

## CONCLUSION

In this study, we provide the mesoscopic model of biological protein materials made of protein crystals based on Go model and virial stress theory. It is shown that our model enables the quantitative predictions of the mechanical properties (e.g. Young's modulus) for biological protein materials, quantitatively and/or qualitatively comparable to AFM experimental result. More remarkably, we suggest the structure-property relation for protein materials such that degree-of-fold, $Q$, representing the folding topology plays a vital role on both Young's modulus and maximum stress exerted to protein materials. For deeper understanding the role of such native topology on mechanical resilience of protein materials, we introduced the simple chain model with $N_B$ hydrogen bonds in two configurations: (i) serial configuration, and (ii) parallel configuration. It is provided that the hydrogen bonds in parallel configuration enhance the mechanical resilience, highlighting the significance of hydrogen bonds in parallel configuration typically observed in β-strand structural motif for mechanical behavior of protein materials. In summary, our model based on Go potential and virial stress theory may make it possible to further understand the structure-property relation for protein materials made of large protein crystal which may be computationally inaccessible with atomistic simulation.

## Acknowledgement

This work was supported in part by KOSEF (Grant No. R01-2007-000-10497-0),

MOST (Grant No. R11-2007-028-00000-0), and BK21 Project (to S.N.) and Nano-Bio Research Center at KIST (to K.E).

**Figure Captions**

Fig. 1. Schematic illustration of biological protein materials composed of protein crystals. (a) cartoon of a fiber, made of protein crystals, under mechanical loading. (b) protein crystal lattices constituting the biological fiber. (c) a unit cell containing a protein crystal

Fig. 2. Stress-strain curves, computed from our mesoscopic model based on Go potential, for biological protein materials composed of model protein crystals

Fig. 3. Stress-strain curve, computed from our mesoscopic model with Tirion's potential, for biological protein materials made of $\alpha$ helix and $\beta$ sheet

Fig. 4. Relationship between degree-of-fold (Q) and contact-order (CO). It is shown that degree-of-fold is highly correlated with contact order such that $Q \approx CO$.

Fig. 5. Relationship between Young's modulus of biological protein materials and degree-of-fold Q. It is shown that degree-of-fold Q is highly correlated with Young's modulus of protein materials

Fig. 6. Relationship between maximum hydrostatic stress of protein materials and degree-of-fold Q. It is provided that degree-of-fold Q is related to the mechanical resilience of protein materials.

Fig. 7. Schematic illustration of a polymer chain with hydrogen bonds (a) in a serial configuration and/or (b) in parallel configuration

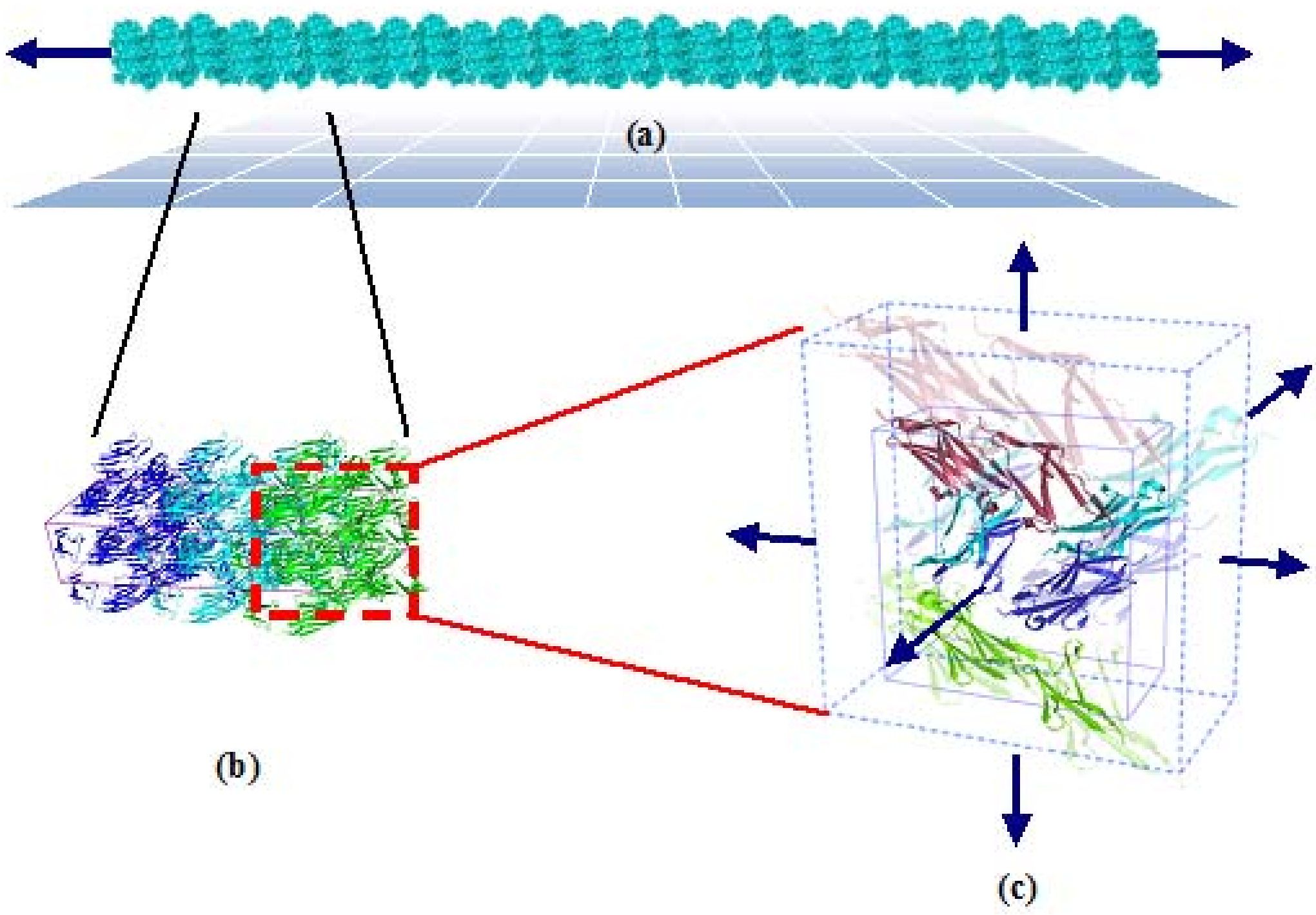

*Fig. 1.* *Yoon, et al*

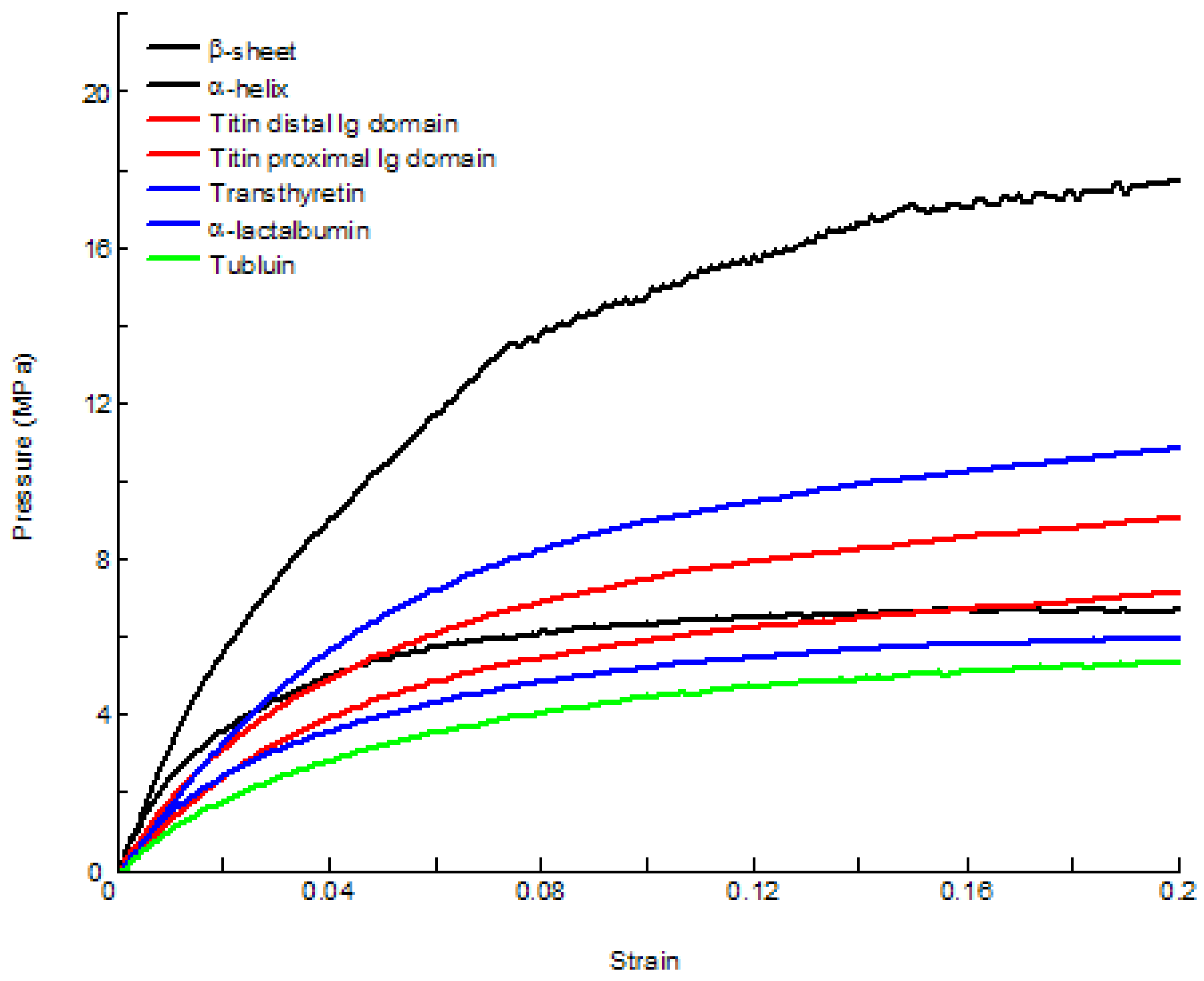

*Fig. 2.* *Yoon, et al.*

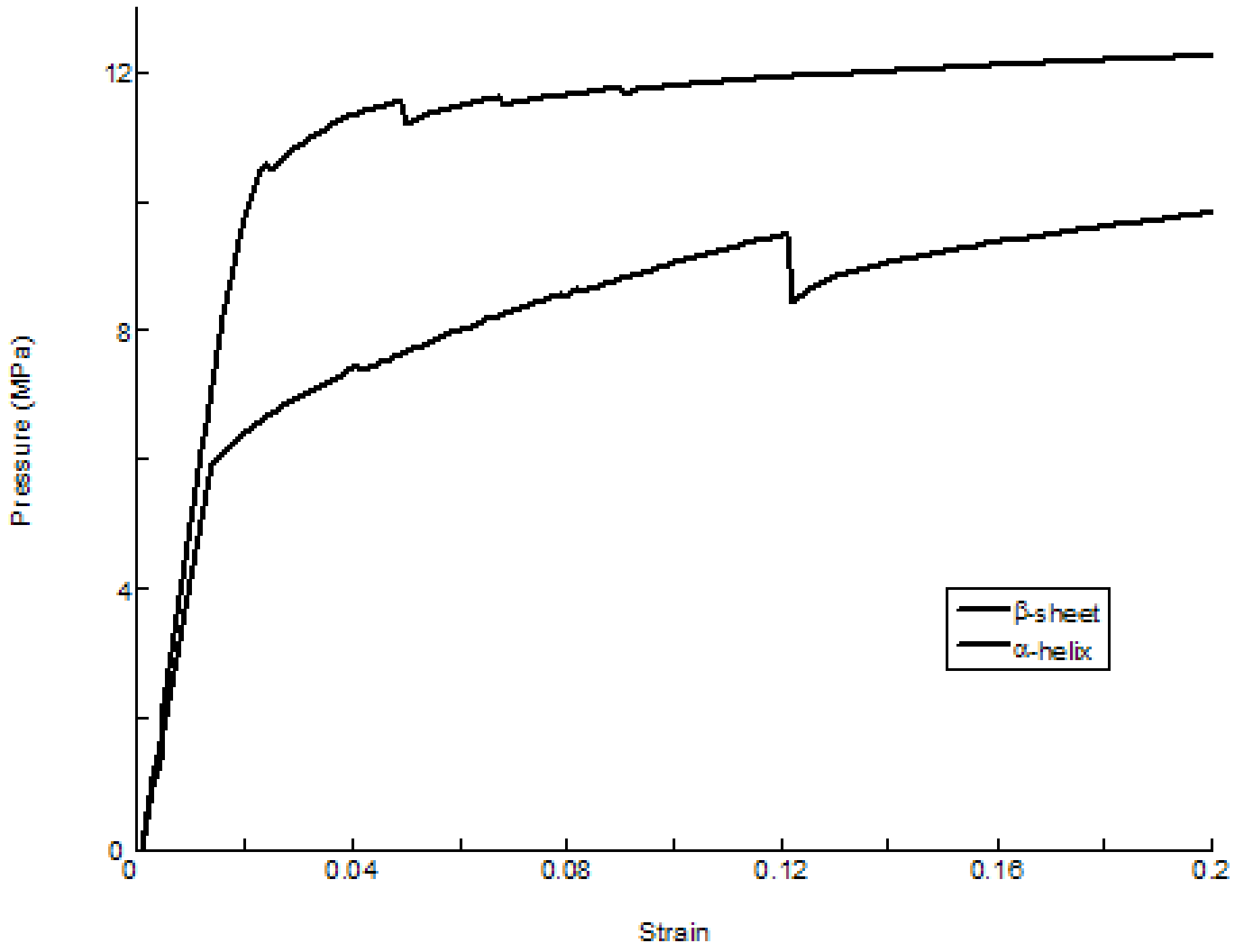

*Fig. 3.* *Yoon, et al.*

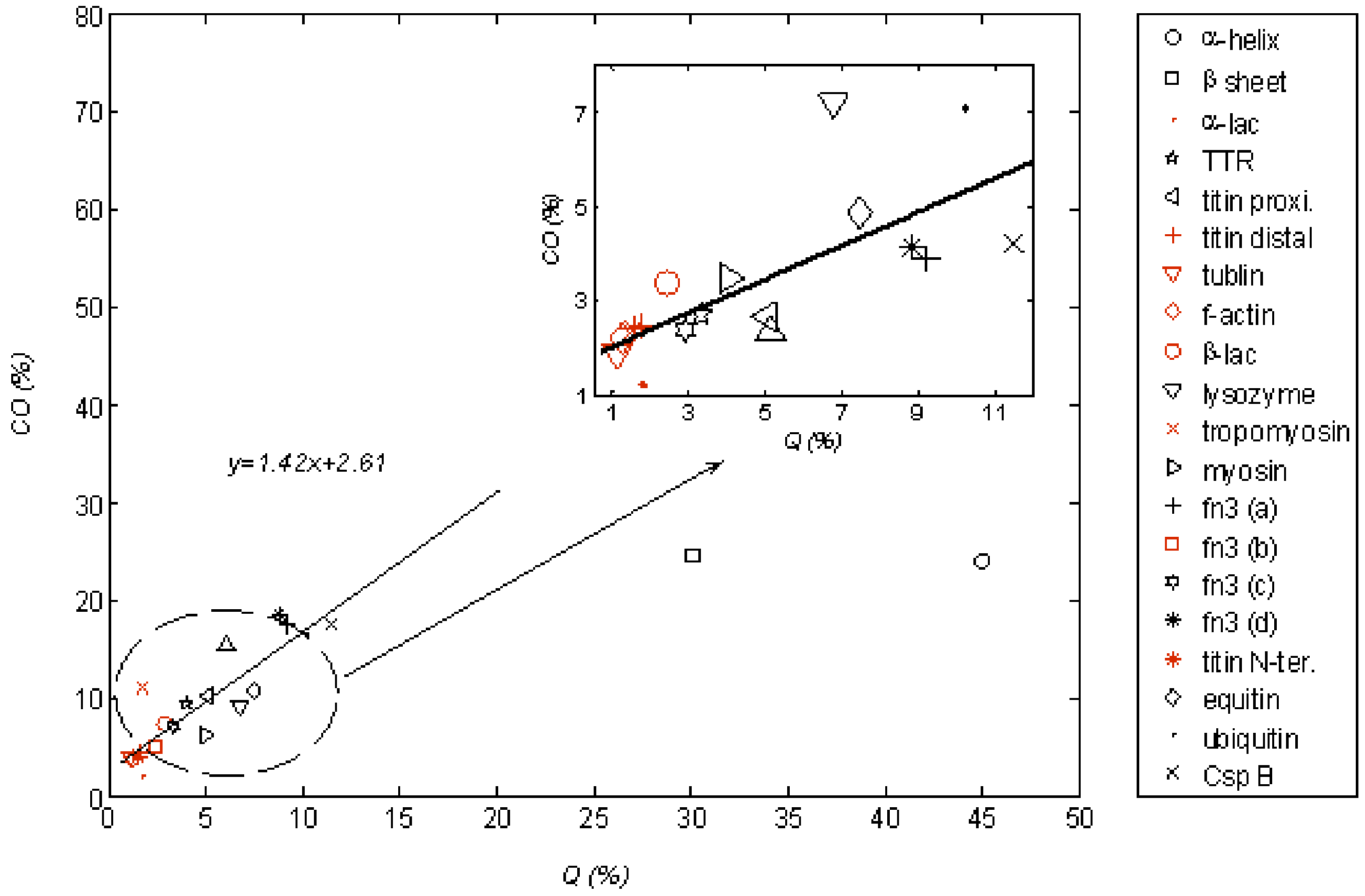

*Fig. 4.* *Yoon, et al.*

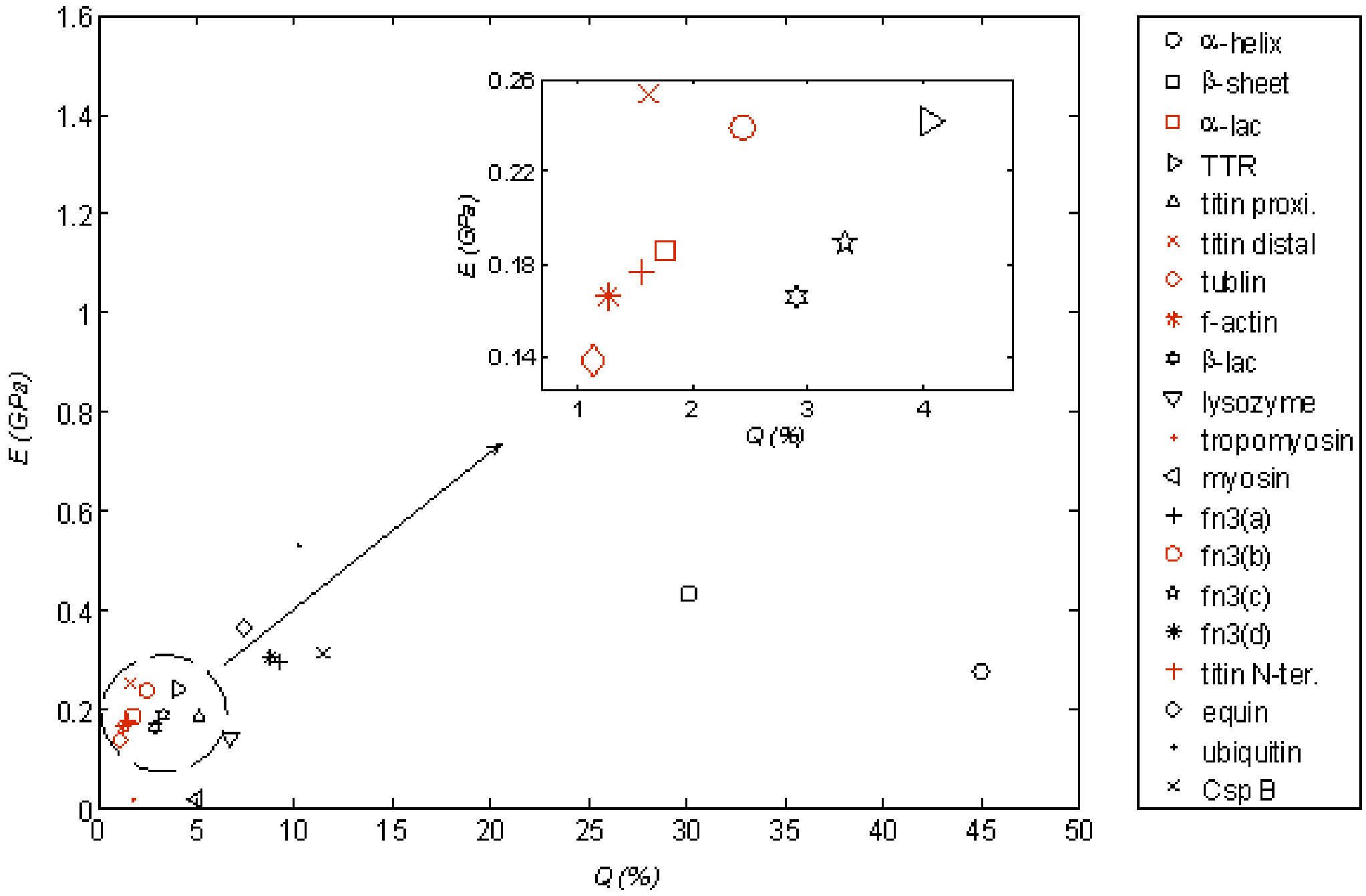

*Fig. 5.* *Yoon, et al.*

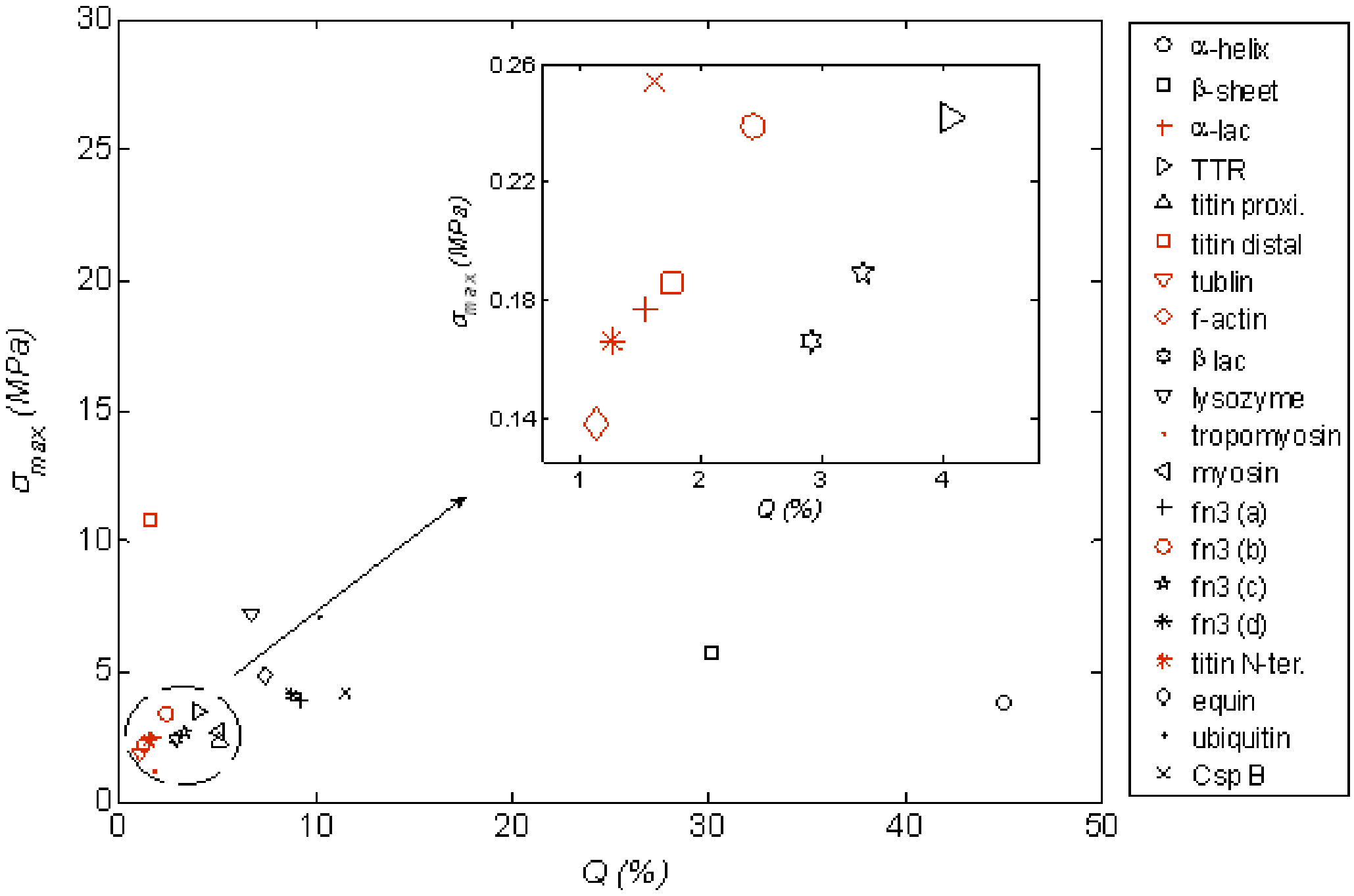

*Fig. 6.* *Yoon, et al.*

(a)

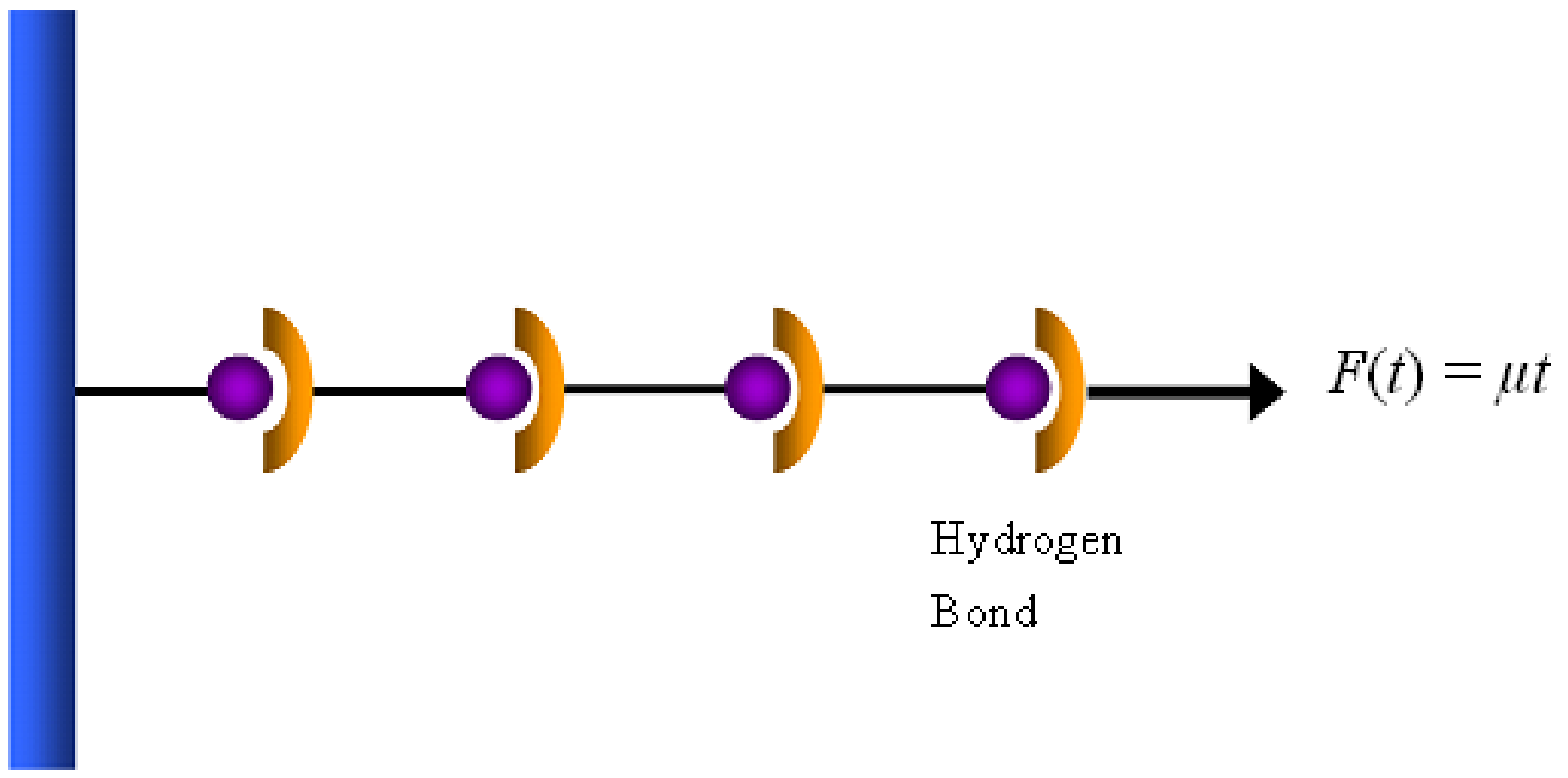

(b)

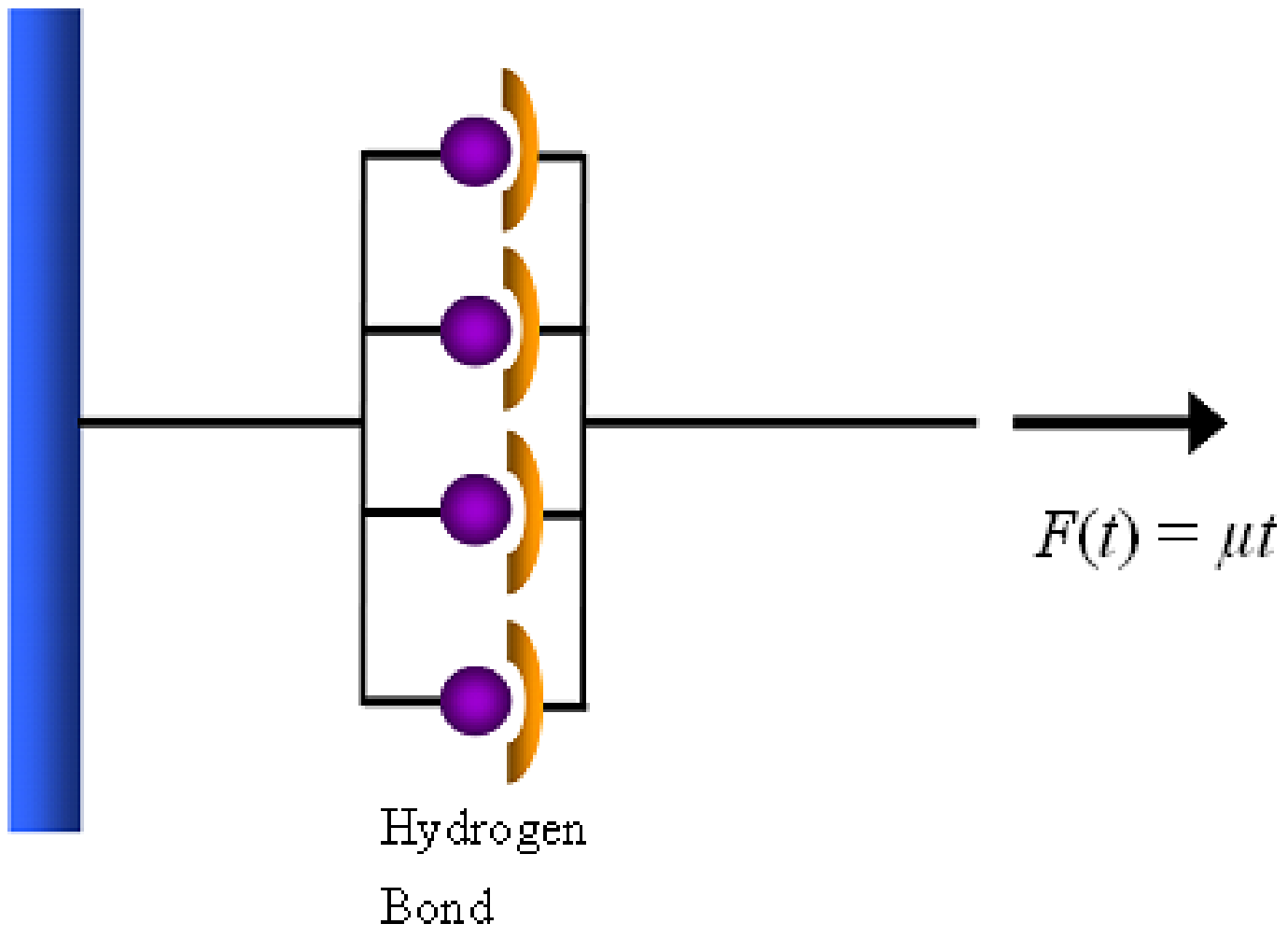

*Fig. 7.* *Yoon, et al.*

**Table 1.** Model Protein Crystals for Biological Protein Materials

| Protein (PDB ID) | Cell parameter | | | Elastic modulus (GPa) | | Maximum stress (MPa) | Q (%) | Contact order (%) |
|---|---|---|---|---|---|---|---|---|
| | Space Group | No. of residue in unit cell | Unit cell volume (Å$^3$) | *in silico*$^*$ | *in vitro*$^\#$ | | | |
| α-helix (1akg) | $P2_12_12_1$ | 64 | 14.6×26.1×29.2 | 0.277 | - | 3.82 | 45.0 | 24.0 |
| β-sheet (2ona) | $P_1$ | 24 | 25.8×9.7×15.8 | 0.433 | - | 5.69 | 30.1 | 24.5 |
| α-lactalbumin (1hfz) | $P2_1$ | 982 | 38.3×78.6×79.6 | 0.186 | 2 | 2.45 | 1.75 | 2.14 |
| Transthyretin (2g5u) | $P2_12_12$ | 908 | 62.2×75.9×134.2 | 0.242 | 5 | 3.48 | 4.05 | 9.34 |
| Titin proximal Ig (1g1c) | $P2_12_12_1$ | 640 | 58.6×60.1×77.1 | 0.187 | - | 2.36 | 5.13 | 10.26 |
| Titin distal Ig (1waa) | $P2_12_12_1$ | 2208 | 43.2×85.8×64.7 | 0.254 | - | 10.8 | 1.61 | 4.44 |
| Tubulin (1tub) | $P2_1$ | 1734 | 80×92×90 | 0.138 | 0.1~2.5 | 1.87 | 1.13 | 3.99 |
| f-Actin rabbit skeletal muscle (1rfq) | $P4_3$ | 2888 | 101.5×101.5×104.2 | 0.166 | 2.2 | 2.19 | 1.26 | 3.86 |
| β-lactoglobulin (1BEB) | $P_1$ | 312 | 37.8×49.5×56.6 | 0.166 | 5 | 2.37 | 2.90 | 7.41 |
| Lysozyme (194L) | $P4_32_12$ | 1032 | 78.65×78.65×37.76 | 0.143 | 5 | 7.21 | 6.77 | 9.16 |
| Fn3 (a) (1fna) | $P2_1$ | 182 | 30.70×35.10×37.70 | 0.294 | - | 3.90 | 9.21 | 17.5 |
| Fn3 (b) (1fnf) | $P2_1$ | 1472 | 64.05×60.67×58.44 | 0.239 | - | 3.36 | 2.43 | 5.10 |
| Fn3 (c) (1fnh) | I2 2 2 | 2152 | 68.58×86.29×142.80 | 0.189 | - | 2.68 | 3.33 | 7.07 |
| FN3 (d) 1ten | $P4_32_12$ | 712 | 49.78×49.78×71.04 | 0.307 | - | 4.12 | 8.81 | 18.6 |
| Titin N-ter (2a38) | P1 | 582 | 55.41×56.29×74.41 | 0.177 | - | 2.40 | 1.54 | 4.05 |
| Equine cyt c (1hrc) | $P4_3$ | 416 | 58.40×58.40×42.09 | 0.365 | - | 4.87 | 7.47 | 10.8 |
| Ubiquitin (1ubq) | $P2_12_12_1$ | 304 | 50.84×42.77×28.95 | 0.532 | - | 7.1 | 10.2 | 16.6 |
| CspB (1csp) | $P3_221$ | 402 | 58.94×58.94×46.45 | 0.315 | - | 4.2 | 11.5 | 17.6 |

$^*$ Young's moduli of model protein materials are computed from our mesoscopic model (*in silico*) based on Go potential field.

$^\#$ Young's moduli of protein fibers are obtained from *in vitro* experiments reported in References[10,51,53].